\documentclass[prd,preprint,showpacs,preprintnumbers,amsmath,amssymb]{revtex4}
\pdfoutput=1

\usepackage{mathrsfs}
\DeclareMathAlphabet{\mathpzc}{OT1}{pzc}{m}{it}

\usepackage[english]{babel}

\usepackage{amsmath}
\usepackage{amssymb}

\usepackage{mathrsfs}
\DeclareMathAlphabet{\mathpzc}{OT1}{pzc}{m}{it}

\usepackage{graphicx}

\usepackage{cancel} % Cro
\usepackage{paralist}
\usepackage{booktabs}
\usepackage{hyperref}
\usepackage{xspace}

\usepackage{graphicx,color,verbatim,acronym,epsfig,subfigure}
\usepackage{slashed,skak}
\usepackage{latexsym,amsfonts,amssymb,amsmath,makeidx,mathrsfs,multirow,lineno,bm,bbm}

\newcommand{\be}{\begin{equation}}
\newcommand{\ee}{\end{equation}}
\newcommand{\ba}{\begin{array}}
\newcommand{\ea}{\end{array}}
\newcommand{\bea}{\begin{eqnarray}}
\newcommand{\eea}{\end{eqnarray}}

\newcommand{\nn}{\nonumber}

\begin{document}
\title{Magnetic field generation from bubble collisions during first-order phase transition}

\author{Jing Yang,}

\author{Ligong Bian}
\email{lgbycl@cqu.edu.cn}
\affiliation{Department of physics, Chongqing University, Chongqing 401331, China}
\begin{abstract}
    We study the magnetic fields generation from the cosmological first-order electroweak phase transition.
We calculate the magnetic field induced by the variation of the Higgs phase for two bubbles and three bubbles collisions.  
Our study shows that electromagnetic currents in the collision direction produce the ring-like magnetic field in the intersect regions of colliding bubbles, which may seed the primordial magnetic field that are constrained by intergalatic field observations.
\end{abstract}
\maketitle

\section{Introduction}

Though the existence of the cosmological magnetic fields has been established by observations, 
its origin is still a long-standing unsolved problem, which may generate during inflation \cite{Turner:1987bw}, and electroweak phase transition \cite{Vachaspati:1991nm}.  
The magnetic fields from the electroweak first-order phase transition (FOPT) may seed the Intergalactic Magnetic Fields\cite{Vachaspati:2001nb}. 
Due to the phase transition in the Standard Model is a {\it cross-over}~\cite{DOnofrio:2014rug}, and the electroweak FOPT is a general prediction of many models beyond the Standard Model, e.g., the SM extended by dimensional-six operator $(\Phi^\dagger \Phi)^3/\Lambda^2$~\cite{Grojean:2004xa,Grojean:2006bp},  singlet extension of the SM~\cite{Profumo:2014opa,Zhou:2019uzq,Zhou:2020idp,Alves:2018jsw,Profumo:2007wc,Espinosa:2011ax,Jiang:2015cwa,Xie:2020wzn,Liu:2021jyc}, two-Higgs-doublet models ~\cite{Cline:2011mm,Dorsch:2013wja,Dorsch:2014qja,Bernon:2017jgv,Andersen:2017ika,Kainulainen:2019kyp}, George-Macheck model~\cite{Zhou:2018zli}, and Next-to minimal supuersymmetry model~\cite{Bian:2017wfv,Huber:2015znp}. Therefore, measurements of Intergalactic Magnetic Fields may provide an additional way to probe physics beyond the Standard Model~\cite{Durrer:2013pga,Vachaspati:2016xji}.
For previous reviews on the primordial magnetic field, we refer to Ref.~\cite{Grasso:2000wj,Durrer:2013pga,Yamazaki:2012pg}. For the status of the observation of magnetic fields in the Galaxy, we refer to Ref.~\cite{jlh}.

 A FOPT proceeds with bubble nucleations and collisions.
 In analogy with the Kibble and Vilenkin~\cite{Kibble:1995aa}, J.Ahonen and K.Enqvist \cite{Ahonen:1997wh} studied the ring-like magnetic fields generation in collisions of bubbles of broken phase in an abelian Higgs model, and evaluated the root-mean-square magnetic field to be around $10^{-21}G$ at the comoving scale of 10 Mpc today after including the turbulent enhancement. T. Stevens et al studied the magnetic field creation from the currents induced by the charged W fields when two bubble collide in Ref.~\cite{Stevens:2007ep}, and they further considered the wall thickness effects in 
 Ref.~\cite{Stevens:2009ty} for two bubble collisons.
 Recently, they utilized the thermal erasure principle to solve the equation of motions (EOMs) of electromagnetic fields in the Non-Abelian Higgs model and found the strength of the magnetic field are comparable to those found in the Abelian Higgs model for two bubbles collision, see Ref.~\cite{Stevens:2012zz}. 
 Different from previous studies, in this work, we take into account the effects of the bubble dynamics during the FOPT, i.e., the dynamics of the bubble walls in the intersecting regions of bubbles and other regions induced by the thermal frictions, see Ref.~\cite{Ellis:2019oqb,Cai:2020djd}. We consider the magnetic field generation by bubble collisions during the electroweak FOPT. For concreteness and simplicity, we consider the magnetic field generation for two- and three- bubble collisions. 
 
This work is organized as follows.
  In Sec.II, we consider the dynamics of general bubble collision. In Sec.III, we solve the EOMs for the W and Z fields, with which, we derive the electromagnetic current by solving the Higgs phase equation and obtain the formula for estimation of the magnetic field. With these preparations, in Sec. IV, we calculate the magnetic field of electroweak bubbles collision in ideal and revised situations
by considering both equal and unequal bubbles collision. In Sec.V, we evaluate the root-mean-squared magnetic field at correlation length after taking into account hydromagnetic turbulent effect. At last, we conclude with Sec.VI.

\section{Bubbles collision dynamics}
\label{bubdny}

At the thin-wall limit, the Lagrangian as a function of the bubble size $R$ can be written as\cite{Darme:2017wvu}:
 \be
L=-4\pi \sigma R^2\sqrt{1-\dot{R}^2}+\frac{4\pi}{3}R^3p\;,
 \ee
where $\sigma$ is the bubble wall tension and $p$ is the pressure acting on the bubble wall. The smallest bubble size of the case
where bubbles would expand 
 instead of collapsing after nucleating is $R=2\sigma /p\equiv R_c$. The EOM to describe bubble growth is given by 
 \be\label{eq:emb}
\ddot{R}+2\frac{1-\dot{R}^2}{R}=\frac{p}{\sigma}(1-\dot{R}^2)^\frac{3}{2}\;.
 \ee
For an expanding bubble, the initial size must be larger than critical radius $R_c$.
We can rewrite Eq.~\ref{eq:emb} in terms of Lorentz factor $\gamma$:
\be\label{eq:embL}
\frac{d\gamma}{dR}+\frac{2\gamma}{R}=\frac{p}{\sigma}\;,
\ee
where $\gamma\equiv 1/\sqrt{1-\dot{R}^2}$. It can be solved analytically by giving an initial condition of $\gamma$ and R.

 When the bubbles are expanding in the plasma background, the friction force can be exerted by the surrounding plasma. In the case where the bubble wall is very relativistic,
the leading-order friction is caused by the change of the effective mass during the $1\rightarrow 1$ particle transmission and reflection in the vicinity of the bubble wall, which is independent of the Lorentz factor $\gamma$, and is estimated to be~\cite{Bodeker:2009qy},
\be
\Delta P_{LO}\approx \frac{\Delta m^2T^2}{24}.
\ee
The next-to-leading order term arising from the particle splitting and transition radiation at the bubble wall is proportional to $\gamma$ \cite{Bodeker:2017cim}:
\be
\gamma \Delta P_{NLO} \approx g^2\Delta m_V T^3,
\ee
where, the squared masses differences between true and false vacuum are given by
 \be
 \Delta m^2\equiv \sum_i c_i N_i \Delta m_i^2 , \ \ \ \ g^2 \Delta m_V\equiv \sum_{i\in V}g_i^2N_i\Delta m_i\;,
 \ee
 where, the sum running over all gauge bosons with its masses changing across the wall, $\Delta m_i=m_{i,t}-m_{i,f}$, $c_i =1(1/2)$ for bosons(fermions),
$N_i$ is the number of internal degrees of freedom of particles, and the $g_i$ are their gauge couplings.
  After these frictions are included, the total pressure can be written as,
 \be
 p\equiv \Delta V - \Delta P_{LO}-\gamma \Delta P_{NLO}\;,
 \ee
 and Eq.~\ref{eq:embL} becomes \cite{Ellis:2019oqb}
 \be\label{eq:embnlo}
 \frac{d\gamma}{dR}+\frac{2\gamma}{R}=\frac{\Delta P_{NLO}}{\sigma}(\gamma_{eq}-\gamma).
 \ee
 where $\gamma_{eq}=(\Delta V - \Delta P_{LO})/(\Delta P_{NLO})$ and $\mathop{lim}\limits_{R\rightarrow \infty}\gamma(R)=\gamma_{eq}$ (We note that the equation is revised in Ref.~\cite{Cai:2020djd} with an additional correction for $\gamma$-dependent friction).  
We assume the Lorentz factor of the bubble wall $\gamma=\gamma_{eq}$ when the bubble collisions take place. 
After the collisions, we have $p\equiv 0$ in the intersection region. Assuming Eqs.~(\ref{eq:embL},\ref{eq:embnlo}) still hold when $p=0$ for the intersection regions of different bubbles, we get
\be
\frac{d\gamma}{dR}+\frac{2\gamma}{R}=0\;,\label{eqR}
\ee
with the solution being
 \be
 \gamma=\gamma_{eq}\frac{R_{col}^2}{R^2}\;,
\ee
 where $R_{col}$ is the radius of the bubble at the collision time $t_{col}$.  And the rest of the bubbles outside the intersection regions are still described by the Eq.~\ref{eq:embL} with $p=\Delta V - \Delta P_{LO}-\gamma \Delta P_{NLO}$, so bubbles still expand with a velocity where $\gamma=\gamma_{eq}$. 
 
  \begin{figure}[!htp]
\includegraphics[scale=0.4]{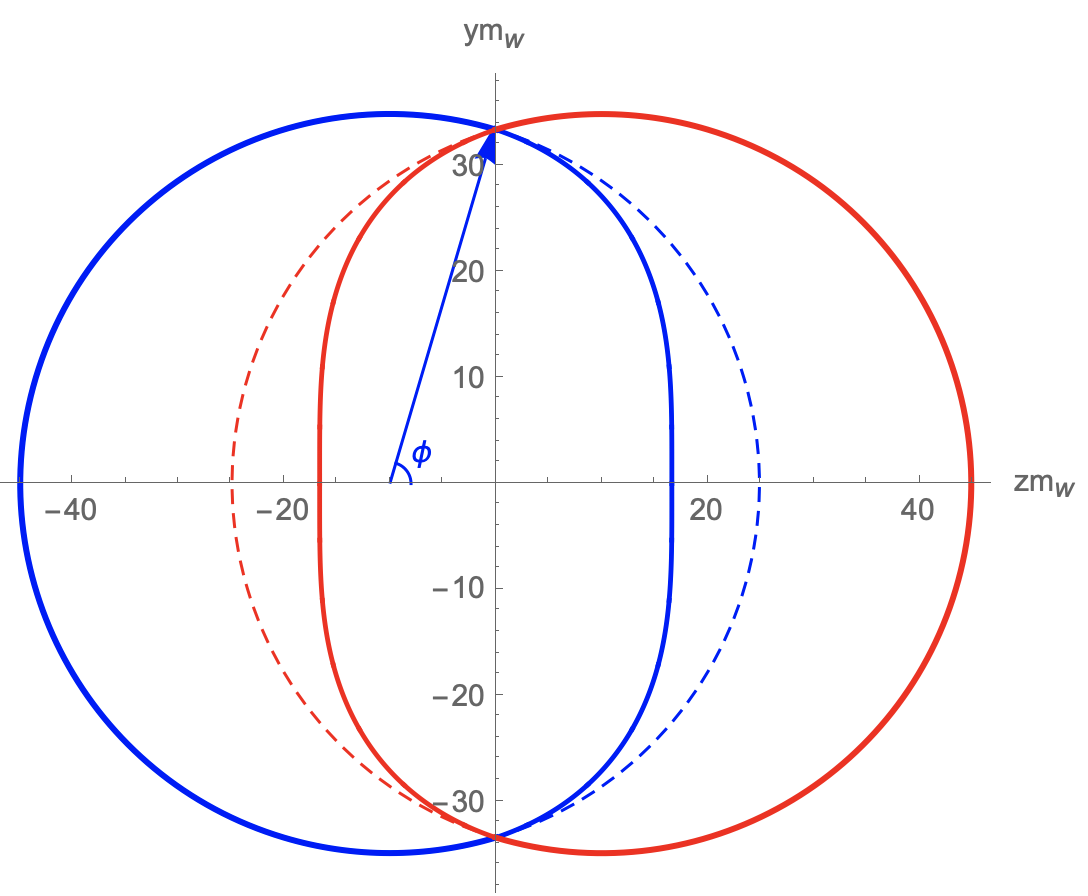}
\includegraphics[scale=0.45]{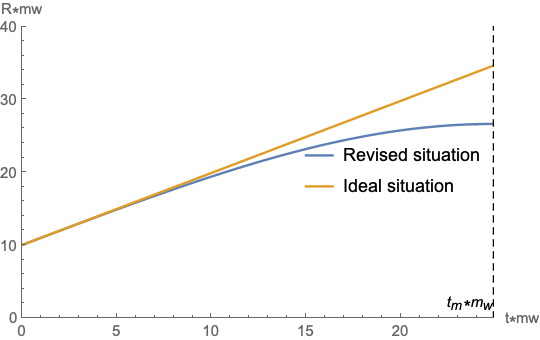}
 \caption{Left: The shape of the two bubbles after collision at time $t'=t_m$ is shown with thick lines, the points on the bubble wall along the collision axis reaches the maximum distance with $R_{col}m_W=10$. 
 The dotted lines are the case where $\gamma \equiv \gamma_{eq}$ is always satisfied on the whole bubble walls. 
 Right: The time dependence of the distance between bubble center and the point of bubble wall along the collision axis in the intersection region is shown in the right panel. The blue line represents the revised situation and the orange one describes the ideal situation.}
 \label{fig2bshape}
\end{figure}
 
 At the time of $t'\equiv t-t_{col}=t_m$, the points on the bubble wall along the collision axis reach the maximum distance away from the bubble center, we show the shape of the bubble walls with solid lines in the left panel of Fig.~\ref{fig2bshape} (which corresponds to the revised situation in the right panel). Where we introduce an angle to describe the position of the points on the wall, the distance between wall and bubble center is related to the angel down from the collision axis, i.e., $\phi$.
  The bubble wall with different $\phi$ begins to intersect with the other bubble at different time. Therefore, the distance between the points on bubble walls in the intersection region and the bubble center depends on the angel ($\phi$) and the time after collision. The Eq.~\ref{eqR} applies until the moment when the points on the bubble wall with a angle $\phi$ reach the maximum distance. Meanwhile, the bubble wall outside the intersection region still expand with a velocity $\gamma=\gamma_{eq}$. The dotted lines are plotted to describe the ideal case which is adopted to evaluate the magnetic field in the previous study of Ref.~\cite{Stevens:2012zz}. In the right panel, we plot the time evolution of bubble walls for revised situation and ideal situation.

\section{Magnetic field generation}

In this section, we review the derivation of the magnetic field from EOMs of gauge bosons. 
  The relevant Lagrangian of electroweak bosonic fields is
\be
L_{EW}=L_1+L_2-V(\phi)\;,
\ee
where, 
\bea
L_1&=&-\frac{1}{4}W_{\mu\nu}^{i}W^{i\mu\nu}-\frac{1}{4}B_{\mu\nu}B^{\mu\nu}\; , \nn \\
W_{\mu\nu}^{i}&=&\partial_\mu W_\nu^i-\partial_\nu W_\mu^i-g\epsilon_{ijk}W_\mu^jW_\nu^k\; , \nn \\
B_{\mu\nu}&=&\partial_\mu B_\nu-\partial_\nu B_\mu\;,
\eea
and
\be
L_2=|(i\partial_\mu-\frac{g}{2}\tau \cdot W_\mu-\frac{g'}{2}B_\mu)\Phi|^2\;,
\ee
where $\tau_i$ is the $SU(2)$ generator and $V(\Phi)$ is the Higgs potential. 
Here, the Higgs potential $V(\phi)$ with proper barrier for quantum tunneling at finite temperature around $\mathcal{O}(10^2)$ GeV can
 feasible an electroweak FOPT proceeding with bubble nucleations and collisions~\cite{Grojean:2004xa}, and therefore yields production of the magnetic fields~\cite{Durrer:2013pga,Kandus:2010nw,Subramanian:2015lua}.
  The physical $Z$ and $A^{em}_\mu$ fields are
\bea
A^{em}_\mu &=& \frac{1}{\sqrt{g^2+g'^2}}(g'W^3_\mu+gB_\mu) \;,\nn \\
Z_\mu &=& \frac{1}{\sqrt{g^2+g'^2}}(gW^3_\mu-g'B_\mu)\;,
\eea
and Higgs doublet takes the form of
\be
\Phi(x)=\left(\ba{ccc} 0 \\
 \rho(x)exp(i\Theta(x))\ea \right)\;,
\ee
where $\Theta(x)$ is the phase of the Higgs field and $\rho(x)$ is its magnitude.
  For this choice of gauge, the EOM for $B$ field is
\be
\partial^2B_\nu-\partial_\nu\partial\cdot B+g'\rho(x)^2\psi_\nu(x)=0\;,
\ee
where the $\psi_\nu$ is
\be
\psi_\nu(x)\equiv \partial_\nu \Theta - \frac{\sqrt{g^2+g'^2}}{2}Z_\nu\;,
\ee
and satisfies
\be\label{eqpsi}
\partial_\nu\left(\rho(x)^2\psi_\nu(x)\right)=0\;.
\ee
For $i=3$, gauge field $W^i$ satisfies the following equation
\be
\partial^2W^3_\nu-\partial_\nu\partial\cdot W^3-g\rho(x)^2\psi_\nu(x)=j^{3}_\nu(x)\;,
\ee
and, for $i=1,2$, we have
\be
\partial^2W^i_\nu-\partial_\nu\partial\cdot W^i+m_W(x)^2W^i_\nu=j^{i}_\nu(x)\;,
\ee
where $m_W(x)^2=g^2\rho(x)^2/2$
and $j^i_\nu(x)$ is,
\bea
j^i_\nu(x) &\equiv& g\epsilon_{ijk}(W^k_\nu\partial\cdot W^j+2W^j\cdot\partial W^k_\nu-W^j_\mu\partial_\nu W^{k\mu}) \nn \\
           &&- g^2\epsilon_{klm}\epsilon_{ijk}W^j_\mu W^{l\mu}W^m_\nu\;.
\eea
  The EOM for $A^{em}$ casts the form of,
\be\label{eqA}
\partial^2A^{em}_\nu-\partial_\nu\partial \cdot A^{em}=j^{em}_\nu(x)\;, \nn \\
\ee
with
\be
j^{em}_\nu(x)=\frac{g'}{\sqrt{g^2+g'^2}}j^3_\nu(x)\;.
\ee
And, the EOM for the Z field is obtained as,
\be\label{eqZ}
\partial^2Z_\nu-\partial_\nu\partial\cdot Z-\rho(x)^2\sqrt{g^2+g'^2}\psi_\nu(x)=\frac{g}{g'}j^{em}_\nu(x)\;.
\ee  
Utilizing the thermal erasure \cite{Stevens:2012zz} of $\langle Z\rangle=0$, and suppose $\rho(x)=\rho_0$, which applies to the thin-wall limit for bubble collisions. Applying the ensemble averaging to Eq.~(\ref{eqpsi},\ref{eqZ}), we get
\bea
\langle j^{em}_\nu\rangle&=&-\frac{g'}{g}\sqrt{g^2+g'^2}\rho_0^2\times \partial_\nu\Theta(x)  \;, \label{eqjem}\\
\partial^2\Theta(x)&=&0\;.\label{eqtheta}
\eea
Consequently, the Eq.~(\ref{eqA}) recasts the form of the Maxwell equation, 
 \bea\label{eqAth}
 \partial^2A_\nu-\partial_\nu\partial\cdot A &=&j^{em}_\nu(x)  \nn \\
 &=&-\frac{g'}{g}\sqrt{g^2+g'^2}\rho_0^2\times \partial_\nu\Theta(x)  \;.
 \eea
Due to the magnetic field 
$
\vec{B}=\vec{\nabla}\times \vec{A}^{em}
$, we can calculate the strength of the magnetic field after obtaining the electromagnetic current through,
\be\label{eqB}
\nabla^2\vec{B}=\vec{\nabla}\times \vec{j}^{em}\;.
\ee

Eq.~\ref{eqjem} suggests that when bubbles collide there would be a large gradient of the Higgs phase, and consequently a 
large electromagnetic current and create large magnetic field through Eq~(\ref{eqAth},\ref{eqB}).
 
 \section{Magnetic field generation}
 
 In this section, we calculate magnetic field generated when two bubbles and three bubbles collide.
 
\subsection{Two Bubbles collision}
   The simplest case is that two bubbles nucleate simultaneously, one bubble locates at $(t,x,y,z)=(0,0,0,vt_{col})$, and the other one locates at the position of $(t,x,y,z)=(0,0,0,-vt_{col})$. We suppose they are expanding with a same velocity $v$,  and thus the collision time is $t_{col}$. The system under study has a $O(2)$ symmetry in the spatial coordinate, we therefore follow the analysis of Kibble and Vilenkin \cite {Kibble:1995aa} and express the EOM in a coordinate $(\tau,z)$ which has a $O(1,2)$ symmetry when $v=1$.  
To obtain the magnetic field generated by bubble wall collisions, we need to solve the equation of the Higgs field phase, i.e., Eq.~\ref{eqtheta}.
In $(\tau,z)$ coordinate, it is 
\be
(\frac{v^2+1}{\tau}+\frac{v^2(1-v^2)t^2}{\tau^3})\frac{\partial\Theta}{\partial\tau}+(1+\frac{v^2(v^2-1)t^2}{\tau^2})\frac{\partial^2\Theta}{\partial\tau^2} \\
-\frac{\partial^2\Theta}{\partial z^2}=0\;, \\
\ee
where $\tau=\sqrt{v^2t^2-r^2}$ with $r^2=x^2+y^2$. Assuming $r\ll vt$, the equation recasts the form:
\be\label{thtauz}
\frac{2}{\tau}\frac{\partial\Theta}{\partial\tau}+v^2\frac{\partial^2\Theta}{\partial\tau^2}-\frac{\partial^2\Theta}{\partial z^2}=0\;.
\ee
 We consider the boundary conditions on $\Theta$ being given by
\be\label{thtaucz}
\Theta(\tau=t_{col},z)=\Theta_0\epsilon(z)\;,\ \frac{\partial}{\partial\tau}\Theta(\tau=t_{col},z)=0\;,
\ee
with $\Theta_0$ being a constant.
  Expressing $\Theta(x)$ as a Fourier transform in z, above equation gives a $\tau-$depend ordinary differential equation, yielding,
\bea\label{eqthz}
\Theta(\tau,z)&=&\frac{1}{\sqrt{2\pi}}\int_{-\infty }^{\infty }dke^{ikz}(a_k\tau^{\frac{-2+v^2}{2v^2}}K_1(\frac{-2+v^2}{2v^2},\frac{\omega_k\tau}{v}) \nn \\
&&+ b_k\tau^{\frac{-2+v^2}{2v^2}}K_2(\frac{-2+v^2}{2v^2},\frac{\omega_k\tau}{v}))\;,
\eea
where $\omega_k=\sqrt{k^2+m^2}$, $a_k$ and $b_k$ are determined by the boundary conditions on $\Theta$. When we take $m\rightarrow 0$, the solution configuration can be obtained.
  Then, the $j_\nu^{em}(\tau,z)$ takes the form
\be
j_\nu^{em}(\tau,z)=(j_z(\tau,z),x_\alpha j(\tau,z))\;,
\ee
with
\bea
 &&j_z = -\frac{g'}{g}\sqrt{g^2+g'^2}\rho_0^2\frac{\partial}{\partial z}\Theta(\tau,z)\;,\\
&&j = -\frac{g'}{g}\sqrt{g^2+g'^2}\rho_0^2\frac{1}{\tau }\frac{\partial}{\partial \tau}\Theta(\tau,z)\;.
\eea
and 
$x_\alpha=(vt,-x,-y)$.
  It is clearly that the electromagnetic field has the same form as the electromagnetic current,
\be
A_\nu^{em}(\tau,z)=(a_z(\tau,z),x_\alpha a(\tau,z))\;.
\ee
Taking the axial gauge, and Maxwell's equation becomes
\be
-\frac{\partial^2}{\partial z^2}a(\tau,z)= j(\tau,z)\;.
\ee
Applying the boundary conditions, namely, $a(\tau_0,z)=0$, and $\partial_z a(\tau=0,z)=0$, we otain
\be
a(\tau,z)=-\int_{-\infty}^z dz'\int_{-\infty}^{z'}j(\tau,z'')dz''\;.
\ee
With which, and apply Eq.~\ref{eqB}, we get the magnetic field,
\bea
B^z &=& 0\;, \nn \\
B^x &=& -y\int_{-\infty }^{z }j(\tau,z')dz'\;, \nn \\
B^y &=& x\int_{-\infty }^{z }j(\tau,z')dz'\;.
\eea

When $v=1$, Eq~\ref{eqthz} reduces to
\be
\Theta(\tau,z)=\frac{\Theta_0}{\tau}\theta(T-|z|)z+\Theta_0\epsilon(z)\theta(|z|-T)\;,
\ee
where $T=\tau-t_{col}$. Then, the $j$ takes the form,
\be
j = \frac{g'}{g}\sqrt{g^2+g'^2}\rho_0^2\frac{\Theta_0}{\tau^3}\theta(T-|z|)z\;.
\ee
Finally, we get
\be
\vec{B}=\frac{(-y,x,0)}{r}B^\phi\;,
\ee
with 
\be\label{eqBphi}
B^\phi=r\frac{g'}{g}\sqrt{g^2+g'^2}\rho_0^2\frac{\Theta_0}{\tau^3}\theta(T-|z|)\times\frac{|z|^2-T^2}{2}\;.
\ee

With increase of the $r$, $\tau$ would decreases, we therefore expect a largest value of the current $j$ and further a largest magnetic field strength at the largest $r$ when two bubbles collide, which grows after the time of bubbles collision ($t_{col}$). This reason leads to a ring-like distribution of the created magnetic field close to the walls of the collided bubbles.

\begin{figure}[!htp]
\includegraphics[scale=0.32]{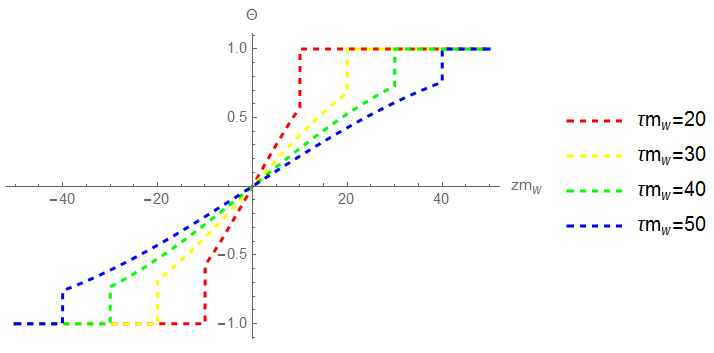} \includegraphics[scale=0.32]{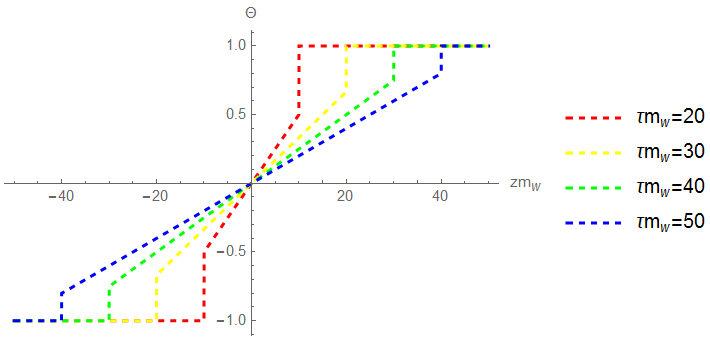}
 \caption{Higgs phase $\Theta$ is shown as a function of the distance z for $\tau m_W=20,30,40,50$, with $\Theta_0=1$. In the left panel, we plot the case of $v=0.5, t_{col} m_W=20$. In the right panel,  we consider the case of $v=1,t_{col} m_W=10$.}
 \label{figtheta}
\end{figure}

\begin{figure}[!htp]
\includegraphics[scale=0.32]{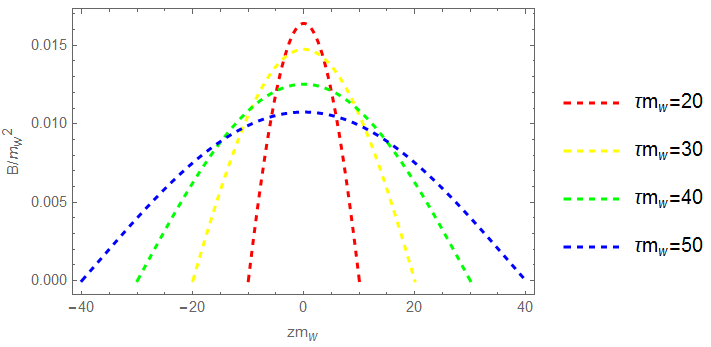} \includegraphics[scale=0.32]{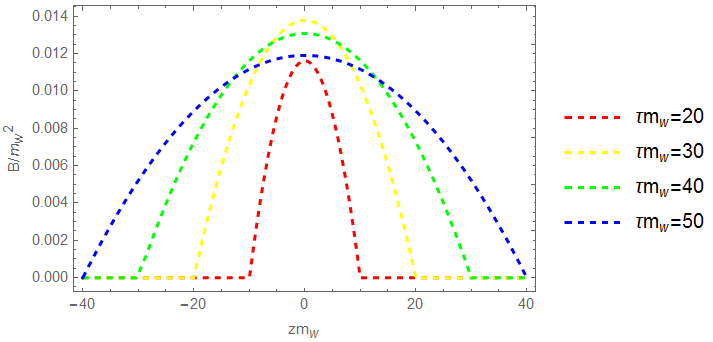}
 \caption{Magnitude of the magnetic field calculated for two bubble collisions. The magnetic field is shown as a function of the distance z along the axis of collision at a distance $rm_W=1$ from the axis of collision for $\tau m_W=20,30,40$ and 50. Left: we consider the case $v=0.5,t_{col} m_W=20$. Right: we consider the case $v=1,t_{col} m_W=10$.}
 \label{figmag2b}
\end{figure}

\begin{figure}[!htp]
\includegraphics[scale=0.50]{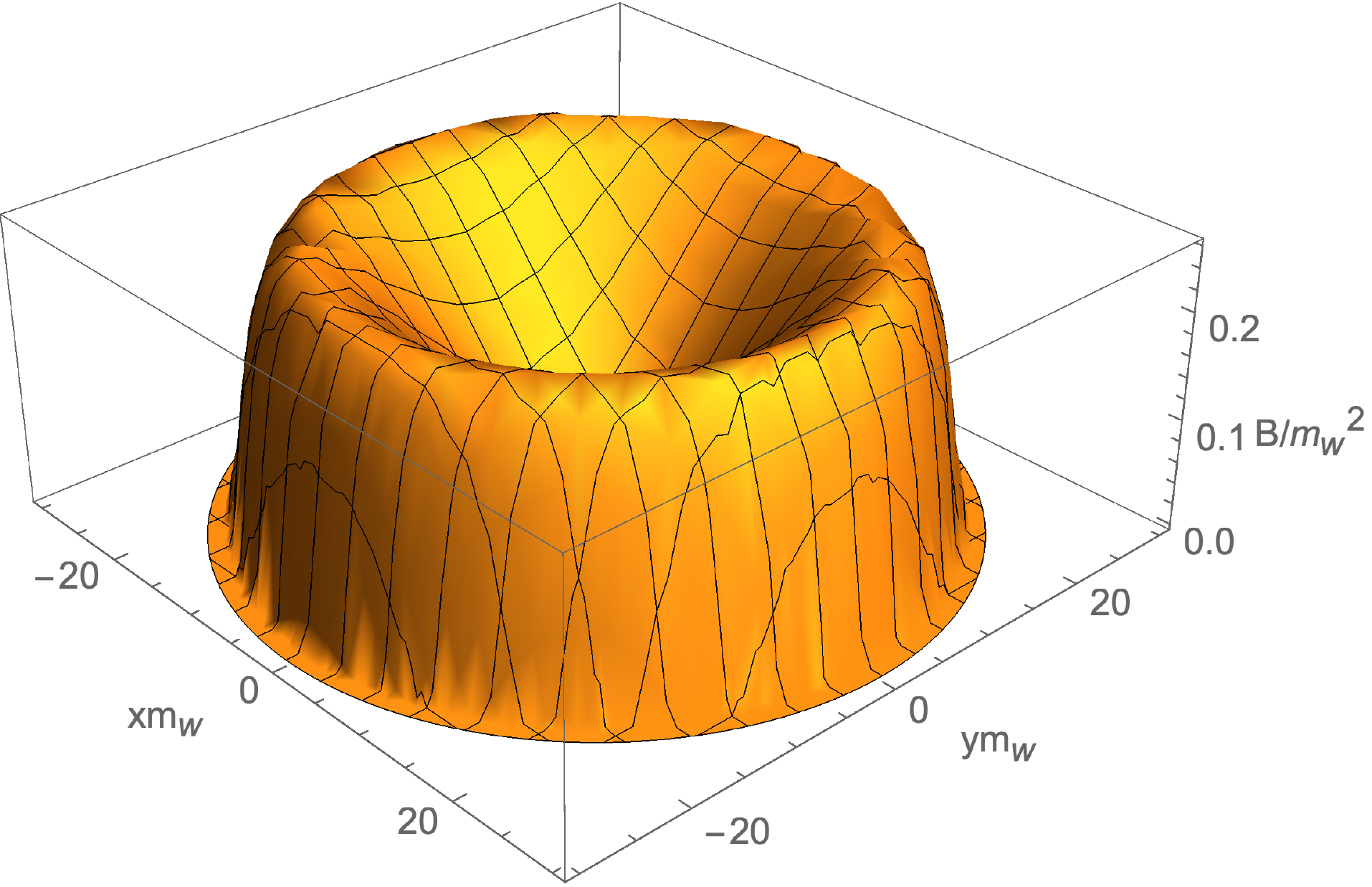}
\caption{Magnitude of the magnetic field calculated for two bubble collisions in the $x-y$ plane for $z=0$ at time $t\;m_W=30$ in the case of $v=1$ and $t_{col}\;m_W=10$.}\label{figB2}
\end{figure}

{\it \bf Equal bubbles-Ideal situation:} We first consider the two colliding bubbles are of equal sizes.
In Fig.~\ref{figtheta}, we show the configuration of the $\Theta$ for $v=0.5$ and $v=1$ as a function of distance z along the axis of collision for different $\tau m_W$. We find there are slightly difference between the cases of $v=0.5$ and $v=1$ for the same $\tau m_W$. Thus the magnetic field strength from the bubble collisions for the two cases have the similar profile as shown in Fig.~\ref{figmag2b}. At a distance $rm_W=1$ and $\tau m_W=20,30,40$ and 50, the magnitude of magnetic is nearly order of $0.01m_W^2$. It can be seen that near the center of the overlap region at $z=r=0$,the magnetic field is much smaller than the region near $\tau=R$, and magnetic field has a tendency to drop at fixed $r$ when the overlap region becomes larger. Magnitude of the magnetic field in $x-y$ plane is shown in Fig.~\ref{figB2}. The figure shows that 
large magnitude of the magnetic field almost distributes near the edge of the overlap region, which indicates that the shape of the magnetic field produced by the electroweak bubble collisions is approximately a ring-like distribution. This feature confirms the discussions under Eq.~\ref{eqBphi}.

  {\it\bf Unequal bubbles-Ideal situation:} Then, we turn to the unequal bubbles collision situation, where the two bubbles nucleate at two different moments. For simplicity, we consider one bubble is nucleated at $(t_1,x_1,y_1,z_1)=(0,0,0,-d_1)$ and the other one at $(t_2,x_2,y_2,z_2)=(d_1-d_2,0,0,d_2)$ where $d_1>d_2>0$. We consider the case where they expand at a velocity $v\equiv 1$ after nucleating, so they would collide at $z=0,t=d_1$. We find nucleation events has a space-like interval due to $\Delta x^\mu \Delta x_\mu=(d_1-d_2)^2-(d_1+d_2)^2<0$. Therefore, one can use an appropriate Lorentz boost to obtain a frame in which the two bubbles are nucleated simultaneously \cite{Kosowsky:1991ua}. In the new frame after the boost, the coordinates $(t'',x'',y'',z'')$ has form
 \be
 t''=\gamma(t-\Delta v \cdot x),\ \ \ x''=x, \ \ \ y''=y, \ \ \ z''=\gamma(z-\Delta v \cdot t)\;,
 \ee
where $\Delta v$ is the velocity of the new frame relative to the old one. The condition that two bubbles nucleate simultaneously requires $t''_1=t''_2$, so we get $\Delta v=(d_1-d_2)/(d_1+d_2)$.

\begin{figure}[!htp]
\includegraphics[scale=0.35]{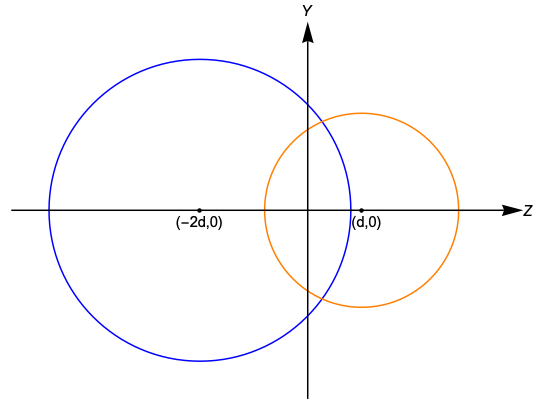} \includegraphics[scale=0.35]{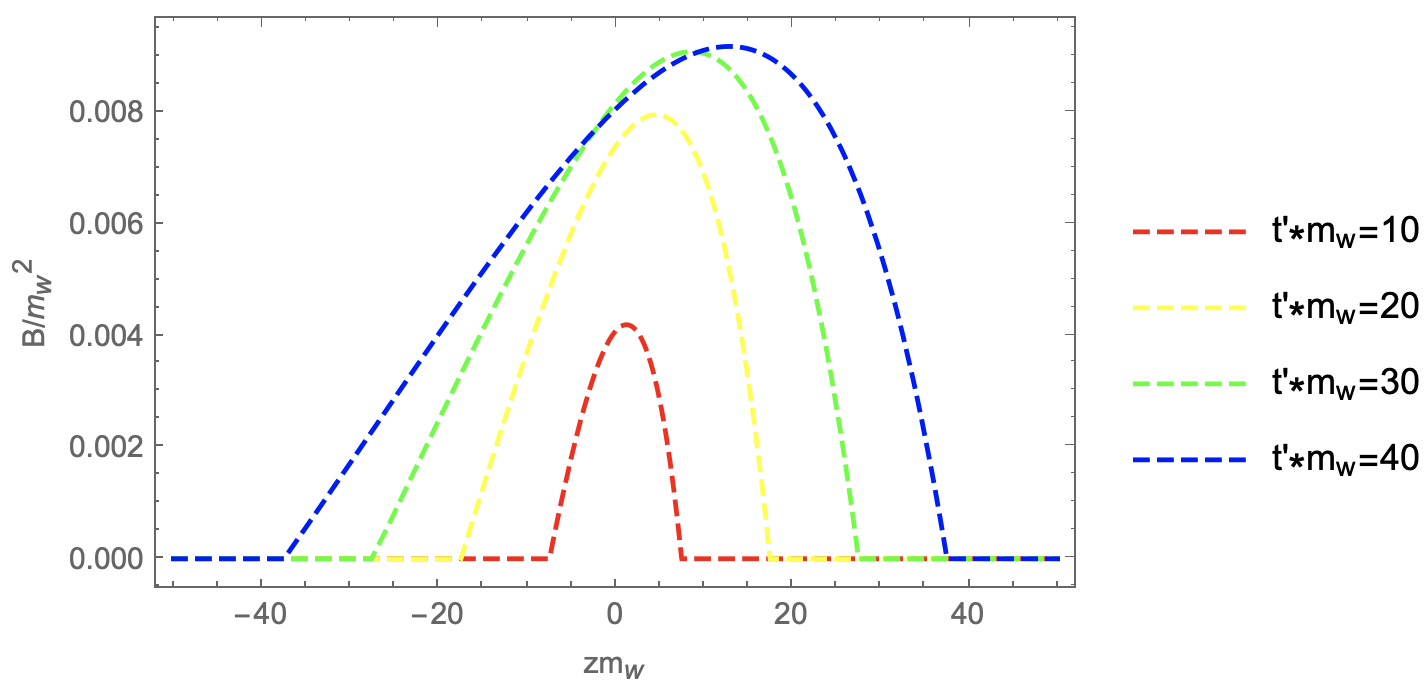}
 \caption{Left: Bubble shapes for unequal bubbles collisions. Right: Magnetic field generated by unequal bubbles collision. The field is shown  at time $t'm_W\equiv (t- t_{col})m_W=10,20,30,40$ after collision, in which we consider $v=1$, $t_{col}m_W=20$,$rm_W=1$.}
 \label{figuneqB}
\end{figure}
     
 We take $d_1m_W=2d_2m_W=20$, and calculate the magnetic field in the new frame using Eq.~\ref{eqBphi}. 
 In order to get the final result, we perform a Lorentz transformation of magnetic field calculated above back to the old frame. 
 Fig.~\ref{figuneqB} shows the bubbles shape for unequal bubbles collision where nucleations occurring at $(t_1,x_1,y_1,z_1)=(0,0,0,-2d)$ and $(t_2,x_2,y_2,z_2)=(d,0,0,d)$ (see the left panel) and the produced magnetic field in the old frame (see the right panel). The Figure shows that magnetic fields for different times are peaked at points with different coordinate z, and shows an asymmetry between left side and right side of peaks since
unequal bubbles collision breaks the $O(1,2)$ and $Z_2$ symmetries in spacetime.

{\it\bf Equal bubbles-Revised situation: }
So far, we have calculated the magnetic field generated by two bubble collisions in an ideal case where the velocity of the whole bubble walls are unchanged after collision and the bubbles are perfect spherical shapes. While, in the realistic situation, the velocity of the intersecting bubble walls may change due to the bubble tension, which lead to a deviation of the bubble shapes, see Section.~\ref{bubdny} for details.
 For a illustration, we suppose that the bubble velocity at the collision time is $v_{col}=v_{eq}=0.99$, and the radius of two bubbles at the collision time are both $R_{col}m_W=10$. To solve the Eq.~\ref{thtauz} by using the boundary conditions Eq.~\ref{thtaucz}, we take an assumption that the solution of $\Theta$ is still nearly proportional to z in the intersection region as shown in Fig.~\ref{figtheta}. It is easily to find that $z/\tau_0$ is a solution to the Eq.~\ref{thtauz} where $\tau_0=\sqrt{t^2-r^2}$. We can approximately take $\tau_{col}=\sqrt{v_{col}^2t^2-r^2}\approx \tau_0$ and the solution takes the form as 
\be\
\Theta(\tau_0,\tau(t,r),z)=C_1(\tau(t,r))*\frac{z}{\tau_0}+C_2\;.
\ee
Then we use the boundary condition Eq.~\ref{thtaucz} and consider the constraint that in the region without intersection one has a constant phase $\pm\Theta_0$. We found the solution is 
\be
\Theta(\tau_0,\tau(t,r),z)=\frac{\Theta_0}{\tau_0}\theta(T(t,r)-|z|)z+\Theta_0\epsilon(z)\theta(|z|-T(t,r))\;,
\ee
where $T(t,r)=\tau(t,r)-R_{col}$ and $\tau(t,r)=\sqrt{R(t,r)^2-r^2}$ with $R(t,r)$ being the distance between bubble wall and bubble center as a function of $r=\sqrt{x^2+y^2}$ and time (t) after collision.

 \begin{figure}[ht]
\includegraphics[scale=0.5]{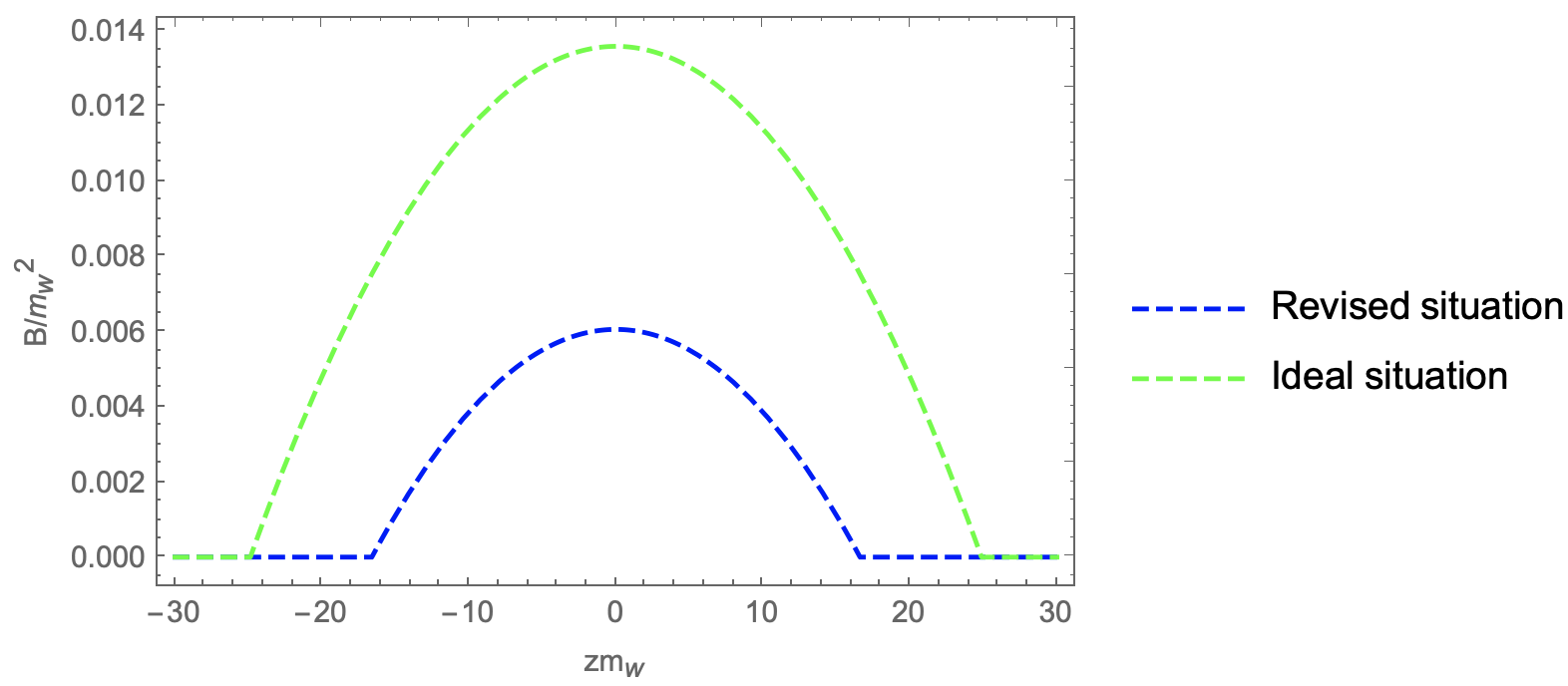}
 \caption{The magnetic field strength is shown as a function of the distance z along the axis of collision with $rm_W=1$ where $v_{col}=0.99,R_{col}m_W=10$. At the time $t-t_{col}=t_m$, the points on bubble walls along the collision axis reaches the maximum distance with bubble center in the revised situation. Blue dashed and Green dashed lines indicate revised and ideal situations respectively. }
 \label{figrev2b}
\end{figure}

  To demonstrate the difference between the revised and ideal situations, in Fig.~\ref{figrev2b}, we show the magnetic field as a function of distance z along the axis of collision with $rm_w=1$. 
We consider the time $t-t_{col}=t_m$ when the points on bubble walls along the collision axis reaches the maximum distance with bubble center in the revised situation. 
The magnitude of magnetic field strength in the revised situation is nearly half of the magnitude in the ideal situation, and the distribution area of magnetic field in the revised situation is smaller than the ideal situation. The results are in accordance with the bubble shape after collision as shown in Fig.~\ref{fig2bshape}, the intersection area of bubbles in the revised situation is smaller than the ideal situation, the electromagnetic current distributes in a smaller area and therefore causes the smaller magnitude of the magnetic field strength. 
    
\subsection{Three bubbles collision}

\begin{figure*}[!htp]
\centerline{\includegraphics[scale=0.37]{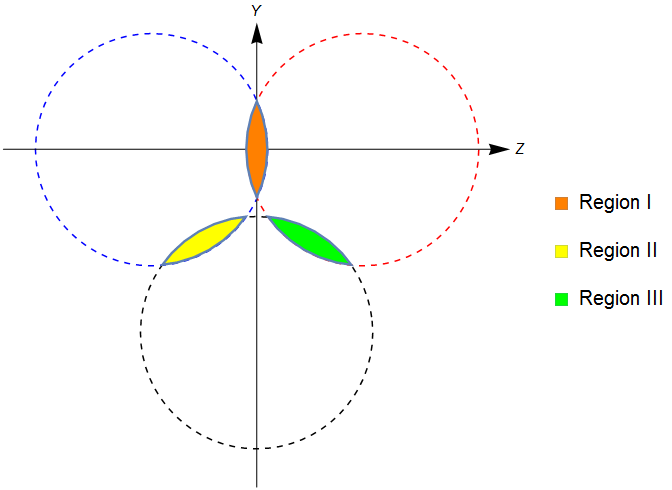}\includegraphics[scale=0.55]{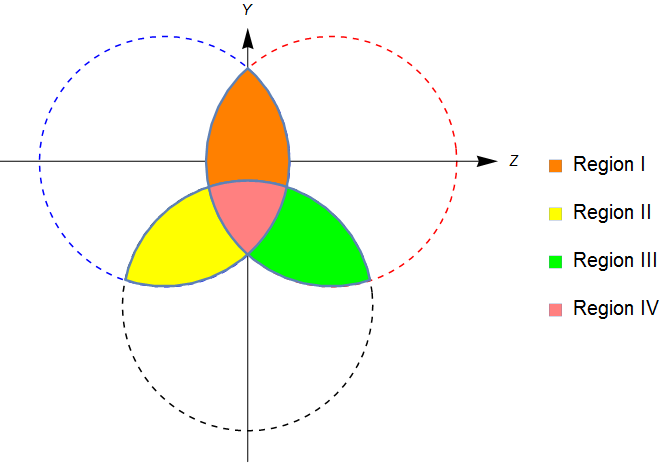}}
\vskip-0.3cm
\caption{Left:overlap regions of three bubbles in the $y-z$ plane at $t-t_{col}=0.1R_{col}$;  Right:overlap regions of three bubbles in the $y-z$ plane at $t-t_{col}=0.5R_{col}$.}
\label{fig3b}
\end{figure*}

  In this section, we consider three equal size bubbles nucleate simultaneously. We consider they expand at a same velocity with $\gamma_{col}=\gamma_{eq}$ and then collide with each other at the same time. The simplest case is that one bubble nucleates at $ (t,x,y,z)=(0,0,0,-R_{col}) $ and other two nucleate at $ (0,0,0,R_{col}) $ and $ (0,0,-\sqrt{3}R_{col},0)$ respectively. At first, there would be three regions where they overlap in pairs. After a period, three regions may overlap and there will be a region (at the center of three bubbles) bounded by the intersection of three bubbles. We show the overlap regions of three bubbles collision in Fig.~\ref{fig3b}. We can imagine that the magnetic field strength of the region IV can be represented by the superposition of the other three regions. For $v_{col}=0.99\approx 1$, we take $\tau=\tau_0=\sqrt{t^2-r^2}$ for simplicity. We set the phases of the three bubbles as $\Theta_1=0$,$\Theta_2=\frac{2\pi}{3}$ and $\Theta_3=\frac{4\pi}{3}$, we choose the center of region I to be the original point, the initial conditions can now be written in the following forms: \\
Region I:
\bea
\Theta(\tau=t_{col},z)&=&\frac{\Theta_1-\Theta_2}{2}\epsilon(z)+\frac{\Theta_1+\Theta_2}{2}, \Theta' (\tau=t_{col},z)=0\;, \nn \\
\tau&=&\sqrt{t^2-x^2-y^2}\;,
\eea
Region II:
\bea
\Theta(\tau'=t_{col},\frac{1}{2}(z-t_{col})-\frac{\sqrt{3}}{2}y)&=&\frac{\Theta_3-\Theta_2}{2}\epsilon(\frac{1}{2}(z-t_{col})-\frac{\sqrt{3}}{2}y) \nn \\
&+&\frac{\Theta_2+\Theta_3}{2} \;,\nn \\
\Theta' (\tau'=t_{col},\frac{1}{2}(z-t_{col})-\frac{\sqrt{3}}{2}y)&=&0\;, \nn
\eea
\be
\tau'=\sqrt{t^2-x^2-(\frac{\sqrt{3}}{2}(z+t_{col})+\frac{1}{2}y)^2}\;,
\ee
Region III:
\bea
\Theta(\tau''=t_{col},-\frac{1}{2}(z+t_{col})-\frac{\sqrt{3}}{2}y)&=&\frac{\Theta_3-2\pi-\Theta_1}{2}\epsilon(-\frac{1}{2}(z+t_{col})-\frac{\sqrt{3}}{2}y) \nn \\
&+&\frac{\Theta_3+2\pi+\Theta_1}{2} \;,\nn \\
\Theta' (\tau''=t_{col},-\frac{1}{2}(z+t_{col})-\frac{\sqrt{3}}{2}y)&=&0 \;,\nn
\eea
\be
\tau''=\sqrt{t^2-x^2-(\frac{\sqrt{3}}{2}(z-t_{col})-\frac{1}{2}y)^2}\;.
\ee

And magnetic field of these three regions can be solved similarly, with: \\
Region I
\bea
B^x &=& -yB_1\;, \nn \\
B^y &=& xB_1\;, \nn \\
B^z &=& 0\;,
\eea
Region II
\bea
B^x &=& -(y-\sqrt{3}z-\sqrt{3} t_{col})\frac{B_2}{2} \;,\nn \\
B^y &=& \frac{x}{2}B_2\;, \nn \\
B^z &=& -\frac{\sqrt{3}}{2} xB_2\;,
\eea
Region III
\bea
B^x &=& -(y+\sqrt{3}z-\sqrt{3} t_{col})\frac{B_3}{2} \;, \nn \\
B^y &=& \frac{x}{2}B_3 \;,\nn \\
B^z &=& \frac{\sqrt{3}}{2}xB_3\;,
\eea
Region IV
\bea
B^x &=& -yB_1-( y-\sqrt{3} z-\sqrt{3} t_{col})\frac{B_2}{2} \;,\nn \\
&-&( y+\sqrt{3} z-\sqrt{3} t_{col})\frac{B_3}{2}\;, \nn
\eea
\bea
B^y &=& xB_1+x\frac{B_2}{2}+x\frac{B_3}{2} \;,\nn \\
B^z &=& \frac{\sqrt{3}}{2} x(B_3-B_2)\;,
\eea
where
\bea
B_1=\frac{g'}{g}\sqrt{g^2+g'^2}\rho_0^2\frac{\Theta_1-\Theta_2}{\tau^2}\theta(\tau-t_{col}-|z|)\frac{|z|^2-(\tau-t_{col})^2}{2} \;,\nn
\eea
\bea
B_2=\frac{g'}{g}\sqrt{g^2+g'^2}\rho_0^2\frac{\Theta_3-\Theta_2}{\tau'^2}\theta(\tau'-t_{col}-|\frac{1}{2}(z-t_{col})-\frac{\sqrt{3}}{2}y|) \nn \\
\frac{|\frac{1}{2}(z-t_{col})-\frac{\sqrt{3}}{2}y|^2-(\tau'-t_{col})^2}{2}\;, \nn
\eea
\bea
B_3=\frac{g'}{g}\sqrt{g^2+g'^2}\rho_0^2\frac{\Theta_3-\Theta_1-2\pi}{\tau''^2}\theta(\tau''-t_{col}-|\frac{1}{2}(z+t_{col})+\frac{\sqrt{3}}{2}y|) \nn \\
\frac{|\frac{1}{2}(z+t_{col})+\frac{\sqrt{3}}{2}y|^2-(\tau''-t_{col})^2}{2}\;. \nn
\eea
\  \  \  \   Note that the solutions are only valid in the overlap region.
  After $t>\frac{2}{\sqrt{3}}t_{col}$, three regions may overlap with each other, and the magnetic field in Region IV is the superposition of the three regions. \\
  
  \begin{figure}[!htp]
\includegraphics[scale=0.4]{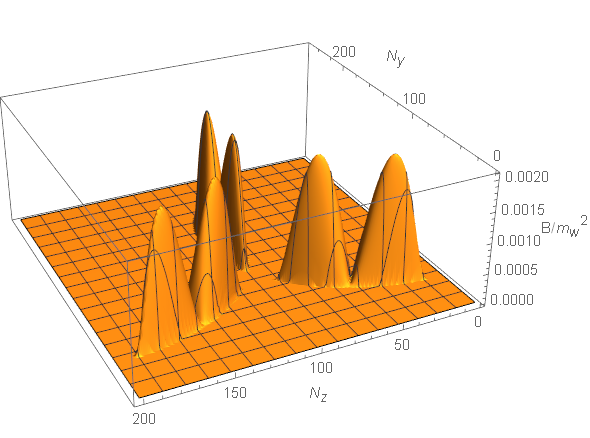}
\includegraphics[scale=0.4]{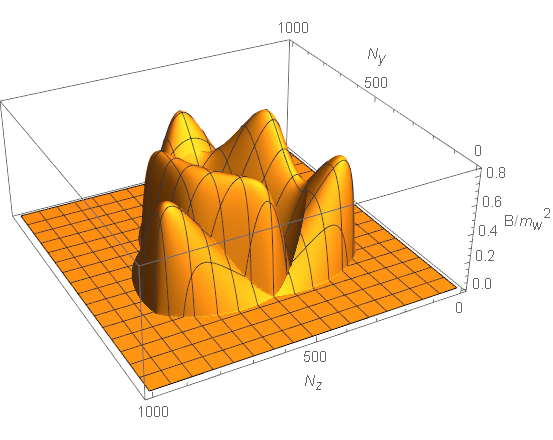}
 \caption{Magnitude of the magnetic field (in the ideal situation) produced by three bubbles collisions for $v_{col}=0.99,R_{col}m_w=t_{col}m_W=10$. We show the field as a function of lattice numbers $N_y$ and $N_z$ on y and z axes respectively with lattice spacing $a=0.1/m_W$.
 Left panel: 
Magnitude of the magnetic field at  $x=0,t-t_{col}=0.1R_{col}$. Right panel: Magnitude of the magnetic field at $x=0,t-t_{col}=t_m$.}\label{fig3bid}
\end{figure}

\begin{figure}[!htp]
\includegraphics[scale=0.4]{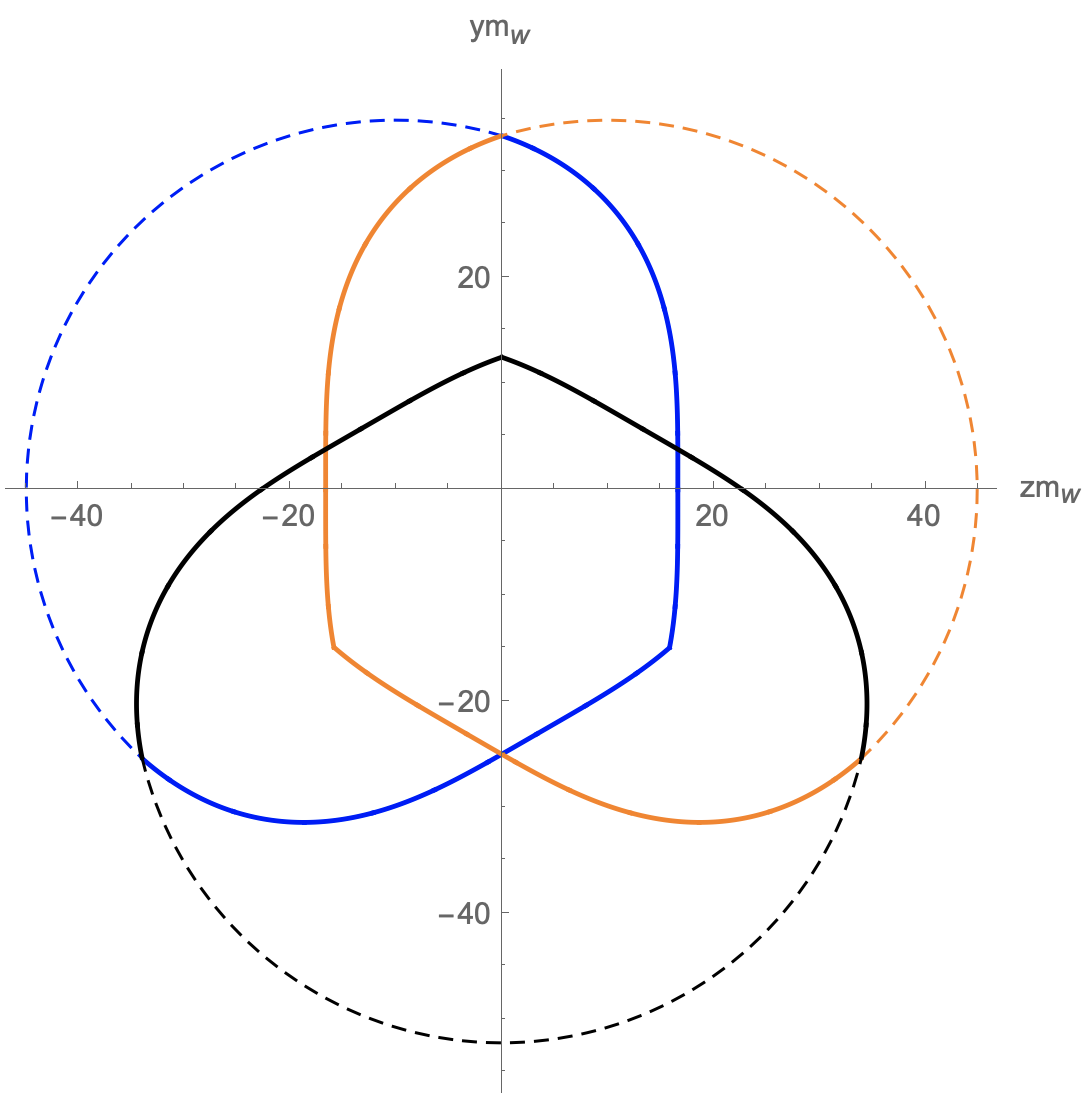}
\includegraphics[scale=0.4]{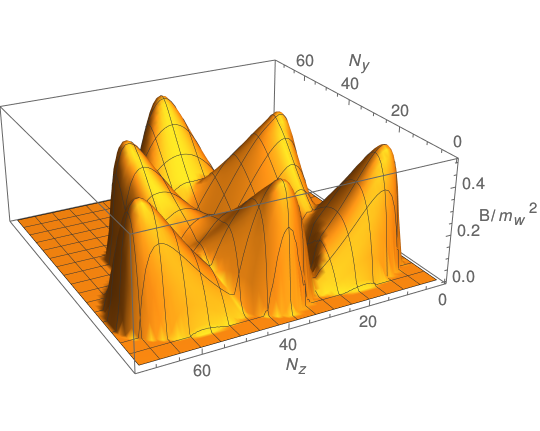}
\caption{Left panel: The shape of three bubbles after collision at $t-t_{col}=t_m$ for $v_{col}=0.99$,$R_{col}m_W=10$. 
Right panel: Magnitude of the magnetic field produced by three bubbles collision. The magnetic field in the revised situation is shown as a function of lattice numbers $N_y$ and $N_z$ on y and z axes respectively with lattice spacing $a=1/m_W$ at $x=0,t-t_{col}=t_m$. }\label{fig3brev}
\end{figure}

 {\it \bf Three equal bubbles-ideal institution:}
 For illustration, we show the strength of the magnetic field induced by three bubbles collision at $t-t_{col}=0.1R_{col}$ and $t-t_{col}=t_m$ in $y-z$ plane in Fig.~\ref{fig3bid}. We can see that the magnetic field distributions of the three regions is separated as expected (see the left panel), where the peak of magnetic field strength is distributed on the symmetric axis of the overlap region.
At a letter time of $t-t_{col}=t_m$, the three regions would overlap and the magnetic field is continuously distributed (see the right panel). The magnitude of the magnetic field is nearly $zero$ in the vicinity of the center of the overlap regions.  And the strength of the magnetic field in region IV has the same order as other three regions.
 
 {\it \bf Three equal bubbles-revised institution:}
     While taking consideration of the real bubble collision situation, the overlap regions of three bubbles collision would be revised as shown in the Fig.~\ref{fig3brev}. The bubbles shape and magnetic field generation of three bubbles collision are calculated at $(x=0, t-t_{col}=t_m)$ with $v_{col}=0.99$ and $R_{col}m_W=10$. In comparison with the ideal situation as shown in Fig.~\ref{fig3bid}, the magnitude of the magnetic field strength in the revised situation is shown to be nearly half of the ideal situation, and the magnetic field distributes more continuously.

\section{Implication for observation}
  The comoving Hubble length at the electroweak phase transition temperature $T_*$ is given by\cite{Kahniashvili:2012uj},
  \be
  \lambda_{H_{*}}=5.8 \times 10^{-10} {\rm Mpc(100 {\rm GeV}/T_{*})} (100/g_{*})^{1/6}\;,
  \ee
  where $g_{*}$ is the number of relativistic degrees of freedom at the moment when the primordial magnetic field is generated. 
The comoving correlation length for a primordial magnetic field at the generation time can be evaluated to be 
 \be
 \xi_\star=\Gamma \lambda_{H_{*}}\;.
 \ee
 Here, $\Gamma$ is the factor to account for bubble numbers inside one Hubble radius at the FOPT, and $\Gamma \simeq 0.01$ for the electroweak FOPT. For recent simulations, see Ref.~\cite{Zhang:2019vsb,Di:2020nny}.

\begin{figure}[!htp]
\includegraphics[scale=0.8]{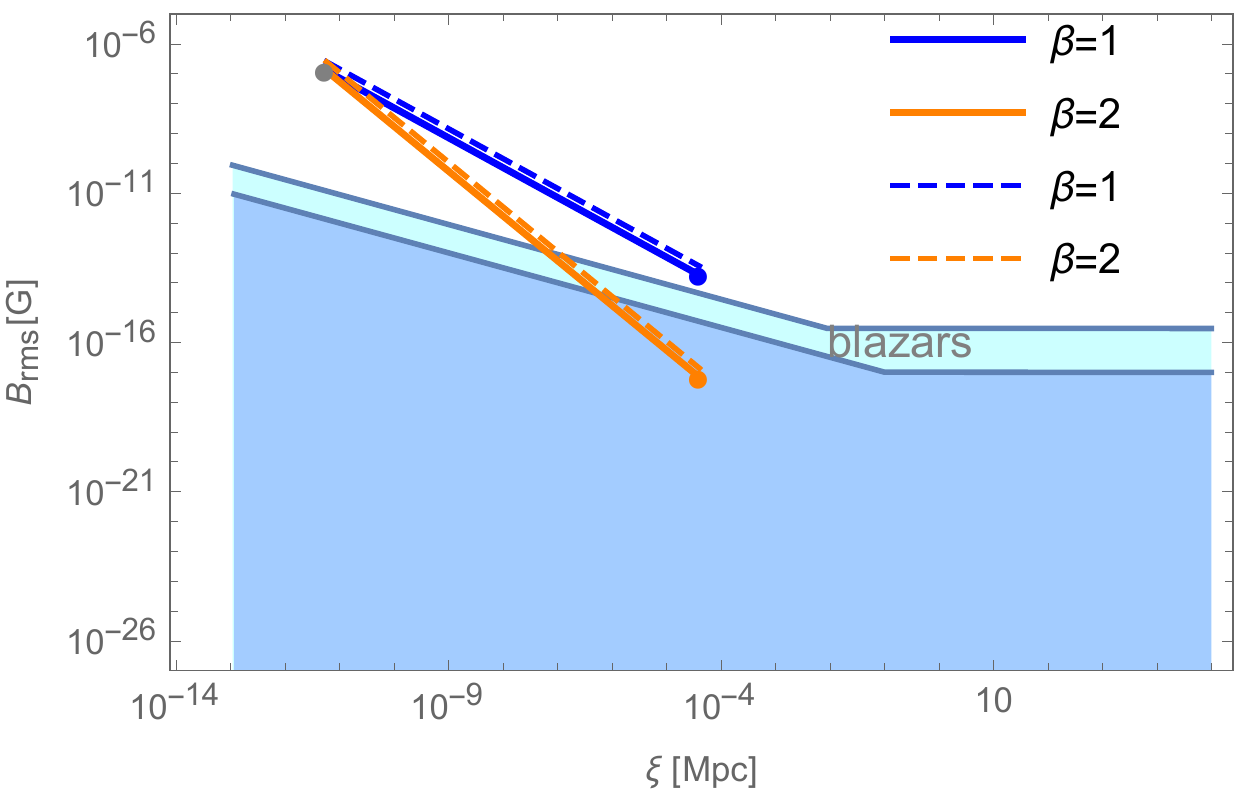}
\caption{ Magetic field strength $B_{rms}$ at the correlation length $\xi$ calculated for the two bubble collision of the {\bf Equal bubbles-Ideal situation} (dashed lines) and {\bf Equal bubbles-Revised situation} (solid lines). Cyan and Blue regions are plotted to consider the bounds set by blazars given in Ref.\cite{Biteau:2018tmv} and Ref.~\cite{Taylor:2011bn}.
 }\label{blazarcons}
\end{figure}

  The physical magnetic field amplitudes scale with the expansion of the universe at the generation time as
 \bea
 B_{*}&=&( \frac{a_{*}}{a_0})^2 B \;,
 \eea
 with the time-temperate relation being,
  \bea
  \frac{a_{*}}{a_0}&\simeq& 8\times 10^{-16}(100{\rm GeV}/T_{*})(100/g_{*})^{1/3}\;.
 \eea
The simulation of the evolution of hydromagnetic turbulence from the electroweak phase transition suggests that the root-mean-squared non-helical magnetic field amplitude and the correlation length satisfies the following relation~\cite{Brandenburg:2017neh},
\be
 B_{rms}=B_{*} (\frac{\xi}{\xi_{*}}) ^{-(\beta+1)/2}\;,
\ee
where $\beta=1,2$ and 4 for non-helical case, with $\xi$ being the magnetic correlation length. For illustration, 
we show in Fig.~\ref{blazarcons} the bounds of blazars spectra on the magnetic field at variant correlation lengths, which depicts that the case of $\beta=1$ case is allowed by the data. Where, we consider the {\bf Equal bubbles-Ideal situation} and the {\bf Equal bubbles-Revised situation} for two bubbles collision, the magnetic fields are generated at $z=0$ with the largest $r$ at $t$($=t_m+t_{col}$), where we have the $r\sim 3 R_{col}$, where we have the ring-like distribution of the magnetic field. The magnetic field strength here is almost the same with the three bubbles collision situations. 
 Primordial magnetic field suffer bounds from the Big-Bang Nucleosynthesis~\cite{Kahniashvili:2010wm,Kawasaki:2012va} and  the measurements of the spectrum and anisotropies of the cosmic microwave background~\cite{Seshadri:2009sy,Ade:2015cva,Jedamzik:1999bm,Barrow:1997mj,Durrer:1999bk,Trivedi:2010gi}, these limits are not shown in the figure since they are not relevant for the parameter space under study in this work.
 
\section{Discussions}
    We use the EOMs of gauge fields to get the magnetic field generated during bubble collisions at the electroweak FOPT. 
    After obtaining the Higgs phases when bubbles collide, we calculate the magnetic field strength after obtaining the electromagnetic current, and apply the approach to the situations of two bubbles and three bubbles collisions, equal and unequal bubbles, ideal and revised situations. We found the electroweak bubble collisions produce the ring-like magnetic field even when we 
consider the revised situation with bubble walls deviating from the spherical shape. For that situation, we get a smaller magnetic field strength because the electromagnetic current distributed in a smaller area. 
The scaling law resulting from the hydromagnetic turbulence after the electroweak FOPT suggests that this kind of magnetic field under study can be probed by the observation of the
Intergalactic Magnetic Fields. The magnetic field strength calculated here is comparable with the magnetic field generated from the bubble collisions simulation performed in Ref.~\cite{Zhang:2019vsb,Di:2020nny}.

 We note that, in the electroweak baryogenesis, the Chern-Simons connects the helicity of the magnetic fields produced during bubble collisions and the baryon asymmetry of the early Universe~\cite{Copi:2008he,Vachaspati:2001nb}. Therefore, the observation of the helicity of the primordial magnetic fields may serve as a test of the electroweak baryogenesis. Ref.~\cite{Ellis:2019tjf} studied the primordial magnetic field from first-order phase transition in $B-L$ model and the SM extended by dimensional-six operator $(\Phi^\dagger \Phi)^3/\Lambda^2$, with the physical implication that the observable gravitational waves and collider signatures would be complementary to the magnetic field observation from the first-order phase transitions.
 when the inverse cascade process are taken into account for helical magnetic fields~
\cite{Cornwall:1997ms,Giovannini:1997gp,Ji:2001rx}.

\section{Acknowledgements}
We thank Yi-Zen Chu, Francesc Ferrer, Jinlin Han, Marek Lewicki, Shao-Jiang Wang, Ke-Pan Xie, and Yiyang Zhang for useful communications and discussions.
This work is supported by the National Natural Science Foundation of China under the grants Nos.12075041, 11605016, and 11947406, and Chongqing Natural Science Foundation (Grants No.cstc2020jcyj-msxmX0814), and the Fundamental Research Funds for the Central Universities of China (No. 2019CDXYWL0029). 

\bibliographystyle{unsrt}
{99}
\end{document}